\documentclass[preprint,prd,nofootinbib,tightenlines,amsmath]{revtex4}
\usepackage{enumerate}
\usepackage{multirow}

\usepackage{epsfig}
\usepackage{graphics,color}      
\usepackage{verbatim}
\usepackage{amsmath}   
\catcode`\@=11
\def\sla@#1#2#3#4#5{{%
 \setbox\z@\hbox{$\m@th#4#5$}%
 \setbox\tw@\hbox{$\m@th#4#1$}%
 \dimen4\wd\ifdim\wd\z@<\wd\tw@\tw@\else\z@\fi
 \dimen@\ht\tw@
 \advance\dimen@-\dp\tw@ \advance\dimen@-\ht\z@
 \advance\dimen@\dp\z@
 \divide\dimen@\tw@ \advance\dimen@-#3\ht\tw@
 \advance\dimen@-#3\dp\tw@ \dimen@ii#2\wd\z@
 \raise-\dimen@\hbox to\dimen4{%
 \hss\kern\dimen@ii\box\tw@\kern-\dimen@ii\hss}%
 \llap{\hbox to\dimen4{\hss\box\z@\hss}}}}

\def\slashed#1{%
 \expandafter\ifx\csname sla@\string#1\endcsname{\rm ~Re}lax
{\mathpalette{\sla@/00}{#1}}
\fi}
\def\declareslashed#1#2#3#4#5{%
 \expandafter\def\csname sla@\string#5\endcsname{%
#1{\mathpalette{\sla@{#2}{#3}{#4}}{#5}}}}
 \catcode`\@=12
\declareslashed{}{/}{.08}{0}{D}
 \declareslashed{}{/}{.1}{0}{A}
 \declareslashed{}{/}{0}{-.05}{k}
 \declareslashed{}{/}{.1}{0}{\partial}
 \declareslashed{}{\not}{-.6}{0}{f}

\def\lsim{\mathrel {\vcenter {\baselineskip 0pt \kern 0pt
    \hbox{$<$} \kern 0pt \hbox{$\sim$} }}}
\def\gsim{\mathrel {\vcenter {\baselineskip 0pt \kern 0pt
    \hbox{$>$} \kern 0pt \hbox{$\sim$} }}}

\begin{document}

\baselineskip=15pt
\preprint{}

\title{Lepton universality violation and right-handed currents in $b \to c \tau \nu$}

\author{Xiao-Gang He$^{1,2,3}$\footnote{Electronic address: hexg@phys.ntu.edu.tw},  German Valencia$^{4}$\footnote{Electronic address: german.valencia@monash.edu }}

\affiliation{$^1$ 
Department of Physics, National Taiwan University, Taipei 10617}

\affiliation{$^2$
Physics Division, National Center for Theoretical Sciences, Hsinchu 30013}

\affiliation{$^3$
Tsung-Dao Lee Institute, and School of Physics and Astronomy, Shanghai Jiao Tong University, Shanghai 200240}

\affiliation{$^4$ School of Physics and Astronomy, Monash University, Melbourne VIC-3800 }

\date{\today}

\vskip 1cm
\begin{abstract}

We consider the recent LHCb result for $B_c\to J/\psi \tau \nu$ in conjunction with the existing anomalies in $R(D)$ and $R(D^\star)$ 
within the framework of a right-handed current with enhanced couplings to the third generation. The model predicts a linear relation between the observables and their SM values in terms of two combinations of parameters. The  strong constraints from $b\to s \gamma$ on  $W-W^\prime$ mixing effectively remove one of the combinations of parameters resulting in an approximate proportionality between all three observables and their SM values. To accommodate the current averages for $R(D)$ and $R(D^\star)$, the $W^\prime$ mass should be near 1 TeV, and possibly accessible to direct searches at the LHC. In this scenario we find that $R(J/\psi)$ is enhanced by about 20\% with respect to its SM value and about 1.5$\sigma$ below the central value of the LHCb measurement. The predicted $d\Gamma/dq^2$ distribution for $B\to D(D^\star) \tau \nu$ is in agreement with the measurement and the model satisfies the constraint from the $B_c$ lifetime.

\end{abstract}

\pacs{PACS numbers: }

\maketitle
    

\newpage

\section{Introduction}

A series of measurements of semileptonic $b\to c \tau \nu$ modes have shown hints of deviations from the standard model (SM) for several years. The mode $B\to D \tau \nu$ has been measured by both BaBar \cite{Lees:2012xj,Lees:2013uzd} and Belle \cite{Huschle:2015rga}; and the mode $B\to D^\star \tau \nu$ has been measured by BaBar  \cite{Lees:2012xj,Lees:2013uzd}, Belle \cite{Huschle:2015rga,Sato:2016svk,Hirose:2016wfn} and LHCb \cite{Aaij:2015yra,LHCbconf}. The average of these measurements performed by the HFLAV \cite{Amhis:2016xyh} collaboration is
\begin{eqnarray}
R(D) &=& \frac{B(\bar{B}\to D\tau^-\bar{\nu}_\tau)}{B(\bar{B}\to D\ell^-\bar{\nu}_\ell)}\, = \, 0.407\pm 0.039\pm 0.024 \nonumber \\
R(D^\star) &=& \frac{B(\bar{B}\to D^\star \tau^-\bar{\nu}_\tau)}{B(\bar{B}\to D^\star\ell^-\bar{\nu}_\ell)}\, = \, 0.304 \pm 0.013 \pm 0.007 \ .
\label{exp}
\end{eqnarray}

The current SM predictions for these quantities are from the lattice for $R(D)$ \cite{Lattice:2015rga,Na:2015kha} and from models for $R(D^\star)$ \cite{Fajfer:2012vx} and are given by
\begin{eqnarray}
R(D) &=& 0.299 \pm 0.011 \nonumber \\
R(D^\star) &=& 0.252 \pm 0.003\ .
\label{smpred}
\end{eqnarray}

Very recently, the corresponding measurement for the mode $B_c^+\to J/\psi \tau^+{\nu}_\tau$ has been reported by LHCb \cite{Aaij:2017tyk}
\begin{eqnarray}
R(J/\psi) &=& \frac{B(B_c^+\to J/\psi \tau^+{\nu}_\tau)}{B(B_c^+\to J/\psi \mu^+{\nu}_\mu)}\, = \, 0.71\pm0.17\pm0.18 \ .
\end{eqnarray}
Different models for the form factors produce a SM result in the range $0.25$ to $0.28$ \cite{Anisimov:1998uk,Kiselev:2002vz,Ivanov:2006ni,Hernandez:2006gt} which is about 2$\sigma$ lower. For definiteness, we will use as SM value the most recent result \cite{Watanabe:2017mip}
\begin{eqnarray}
R(J/\psi)=0.283\pm0.048\ .
\end{eqnarray}

Not surprisingly, these anomalies have generated a large number of possible new physics explanations including additional Higgs doublets, gauge bosons and leptoquarks \cite{Kamenik:2008tj,Tanaka:2010se,Crivellin:2012ye,Celis:2012dk,Deshpande:2012rr,Fajfer:2012jt,Datta:2012qk,Becirevic:2012jf,He:2012zp,Rashed:2012bd,Sakaki:2012ft,Tanaka:2012nw,Ko:2012sv,Bambhaniya:2013wza,Atoui:2013zza,Dutta:2013qaa,Freytsis:2015qca,Boucenna:2016wpr,Boucenna:2016qad,Chiang:2016qov,Zhu:2016xdg,Kim:2016yth,Celis:2016azn,Dutta:2017xmj,Cvetic:2017gkt,Ko:2017lzd,Chen:2017hir,Chen:2017eby,Dutta:2017wpq}.

In Ref.~\cite{He:2012zp} we have studied $R(D)$ and $R(D^\star)$ in the context of a right-handed $W^\prime$ with enhanced couplings to the third generation \cite{He:2002ha,He:2003qv}. Here, we revisit this possibility motivated by the new measurement of $R(J/\psi)$, and to address additional constraints from the $d\Gamma/dq^2$ distributions \cite{Freytsis:2015qca} and the $B_c^\pm$ lifetime \cite{Alonso:2016oyd}.

\section{Charged current interactions}

The gauge group of our model is $SU(3)_C\times SU(2)_L\times SU(2)_R \times U(1)_{B-L}$, but the three generations of fermions are chosen to transform differently to achieve non-unversality. In the weak interaction basis, the first two generations of quarks $Q_L^{1,2}$, $U_R^{1,2}$, $D_R^{1,2}$ transform as $(3,2,1)(1/3)$, $(3,1,1)(4/3)$ and $(3,1,1)(-2/3)$, and the leptons $L_L^{1,2}$, $E_R^{1,2}$ transform as $(1,2,1)(-1)$ and $(1,1,1)(-2)$. The third generation, on the other hand, has transformation properties under the gauge group given by  $Q_L^3\;(3,2,1)(1/3)$, $Q^3_R\;(3,1,2)(1/3)$, $L^3_L\;(1,2,1)(-1)$ and $L^3_R\;(1,1,2)(-1)$. In this way $SU(2)_R$  acts only on the third generation and singles it out providing the source of lepton universality violation to explain the anomalies mentioned above. The model is detailed in Refs.~\cite{He:2002ha,He:2003qv}, but here we provide its salient ingredients.

To separate the symmetry breaking scales of $SU(2)_L$ and $SU(2)_R$, we introduce the two 
Higgs multiplets $H_L\; (1,2,1)(-1)$ and $H_R\;(1,1,2)(-1)$ with respective vevs $v_L$ and $v_R$. An additional bi-doublet $\phi\;(1,2,2)(0)$ scalar with vevs $v_{1,2}$ is  needed to provide mass to the fermions. Since both $v_1$ and $v_2$ are required to be non-zero for fermion mass generation, the  $W_L$ and $W_R$ gauge bosons of $S(2)_L$ and $SU(2)_R$ will mix with each other. In terms of the mass eigenstates $W$ and $W'$, the mixing can be parametrized as
\begin{eqnarray}
W_L &=& \cos\xi_W W - \sin \xi_W W'\;,\nonumber\\
W_R &=& \sin\xi_W W +\cos\xi_W W'\;.
\end{eqnarray}
In the mass eigenstate basis the quark-gauge-boson interactions are 
given by,
\begin{eqnarray}
{\cal L}_W&=& - \frac{g_L}{ \sqrt{2}} \bar U_L \gamma^\mu V_{KM} D_L
(\cos\xi_W W^{+}_\mu - \sin\xi_W W^{'+}_\mu)\nonumber\\
&&-\frac{g_R}{ \sqrt{2}}
\bar U_{R} \gamma^\mu V_{R} D_{R}
(\sin\xi_W W^{+}_\mu + \cos\xi_W W^{'+}_{\mu}) ~+~{\rm h.~c.},
\label{cccoup}
\end{eqnarray}
where $U = (u,\;\;c,\;\;t)$, $D = (d,\;\;s,\;\;b)$, $V_{KM}$ is
the Kobayashi-Maskawa mixing matrix and $V_R \equiv (V_{Rij})=(V^{u*}_{Rti}V^{d}_{Rbj})$ with $V^{u,d}_{Rij}$ the unitary matrices
which rotate the right handed quarks $u_{Ri}$ and $d_{Ri}$ from the weak eigenstate basis 
to the mass eigenstate basis. 

The model has a different neutrino spectrum than the SM: three left-handed neutrinos $\nu_{L_i}$ and one right-handed neutrino $\nu_{R_3}$.
Additional scalars $\Delta_L\;(1,3,1)(2)$ and $\Delta_R\;(1,1,3)(2)$ with vevs $v^{L,R}_\Delta$ are needed to generate neutrino masses. In order for the  possibly enhanced $SU(2)_R$ interaction with the third generation to explain the $B$ decay anomalies, we  need the right-handed neutrino 
to be light, which requires $v^{L,R}_\Delta$ to be small. In this model, the neutrinos will receive Majorana masses from the vevs of $\Delta_{L,R}$ and Dirac masses from $\phi$. The mass eigenstates $(\nu^m_L, (\nu^m_{R_3})^c)$  
 are related by a unitary transformation to the weak eigenstates as
\begin{eqnarray}
\left (\begin{array}{c}
\nu_L\\
\nu^c_{R_3}
\end{array}
\right )
= \left ( \begin{array}{cc}
U_L&U_{RL}\\
U_{LR}&U_R
\end{array}
\right )
\left (\begin{array}{c}
\nu^m_L\\
(\nu^m_{R_3})^c
\end{array}
\right )\;.
\end{eqnarray}
In our model $U_L= (U_{Lij})$, $U_{RL} = (U_{RLi3})$ and $U_{LR} = (U_{LR3i})$ and $U_R = (U_{R33})$ are
$3\times 3$, $3\times 1$, $1\times 3$ and $1\times 1$ matrices, respectively. 

Writing the rotation of charged lepton weak eigenstates $\ell_{L,R}$ into mass eigenstates $\ell^m_{L,R}$ as $\ell_{L,R} = V^\ell_{L,R} \ell^m_{L,R}$, the lepton interaction with $W$ and $W'$ becomes
\begin{eqnarray}
{\cal L}_W &=& - \frac{g_L}{ \sqrt{2}} (\bar \nu_L \gamma^\mu U^{\ell \dagger}\ell_L 
+ \bar \nu_{R3}^c \gamma^\mu U^{\ell *}_{RLj3}\ell_{Lj})
(\cos\xi_W W^{+}_\mu - \sin\xi_W W^{'+}_\mu) \nonumber \\
&-&\frac{g_R}{ \sqrt{2}}
(\bar \nu^c_{Li} \gamma^\mu  U^\ell_{LRij} \ell_{Rj} + \bar \nu_{R3} \gamma^\mu U^\ell_{R3j}\ell_{Rj})
(\sin\xi_W W^{+}_\mu + \cos\xi_W W^{'+}_{\mu})  ~+~{\rm h.~c.},
\label{cccouplep}
\end{eqnarray}
where
\begin{eqnarray}
U^{\ell\dagger} = U^{ \dagger}_L V^\ell_L\;,\;\;
U^{\ell *}_{RLj3} = (U_{RLi3}^{*}V^\ell_{Lij})\;,\;\;
U^\ell_{LRij} = U_{LR3i} V^\ell_{R3j}\;,\;\;
U^\ell_{R3j}=U_{R33} V^\ell_{R3j}\;.
\end{eqnarray}
$U^\ell$ is approximately the PMNS matrix.
From Eqs.~\ref{cccoup} and \ref{cccouplep} we see that a large $g_R/g_L$ will enhance the third generation interactions with $W^\prime$.

The final neutrino flavor is not identified in $B$ meson decays so it must be summed.  For the processes involving left- and right-handed charged leptons, neglecting neutrino masses compared with the charged lepton masses, the final decay rates into a charged lepton $\ell_j$, when summed over the different neutrino final states, are proportional to 
\begin{eqnarray}
\mbox{For $\ell_{Lj}$}: &&\sum_i |U^\ell_{ij}|^2 +|U^\ell_{RLj3}|^2  = \sum_i |U^*_{Lli}V^\ell_{Llj}|^2 +|U^*_{RLl3}V^\ell_{Llj}|^2\nonumber\\
&&=(\sum_i |U^*_{Lli}|^2 +|U^*_{RLl3}|^2)|V^\ell_{Llj}|^2= \sum_l |V^\ell_{Llj}|^2 = 1\;,\nonumber\\
\mbox{For $\ell_{Rj}$}: &&\sum_i |U^\ell_{LRij}|^2 +|U^\ell_{R3j}|^2  = \sum_i |U_{LR3i}V^\ell_{R3j}|^2 +|U_{R33}V^\ell_{R3j}|^2\nonumber\\
&&=(\sum_i |U_{LR3i}|^2 +|U_{R33}|^2)|V^\ell_{R3j}|^2 =|V^\ell_{R3j}|^2\;.
\end{eqnarray}
To obtain these results we used the unitarity of $U$: $\sum_i |U^*_{L\ell i}|^2 +|U^*_{RL\ell 3}|^2=1$, 
$\sum_i |U_{LR3i}|^2 +|U_{R33}|^2=1$ and the unitarity of $V_{L,R}^\ell$.

\section{$\Gamma$ and $d\Gamma/dq^2$ predictions}

The starting point for our calculations is the differential decay rate $d\Gamma/dq^2$ with $q^2=(p_B-p_{D^{(\star)}})^2$ for the SM. We use the notation, parameterization and values of Ref.~\cite{Lees:2013uzd} for all the relevant form factors.  This will be sufficient for a comparison to the experimentally determined shape of this distribution as well as the total decay rate.  Of course, $d\Gamma/dq^2$ is obtained after integrating over angles and summing over polarizations. Other observables, such as angular correlations, can also be used to discriminate between the SM and new physics scenarios as well, but we will not consider that possibility in this paper as they have not been measured  yet.

\begin{figure}[!htb]
\begin{center}
\includegraphics[width=6cm]{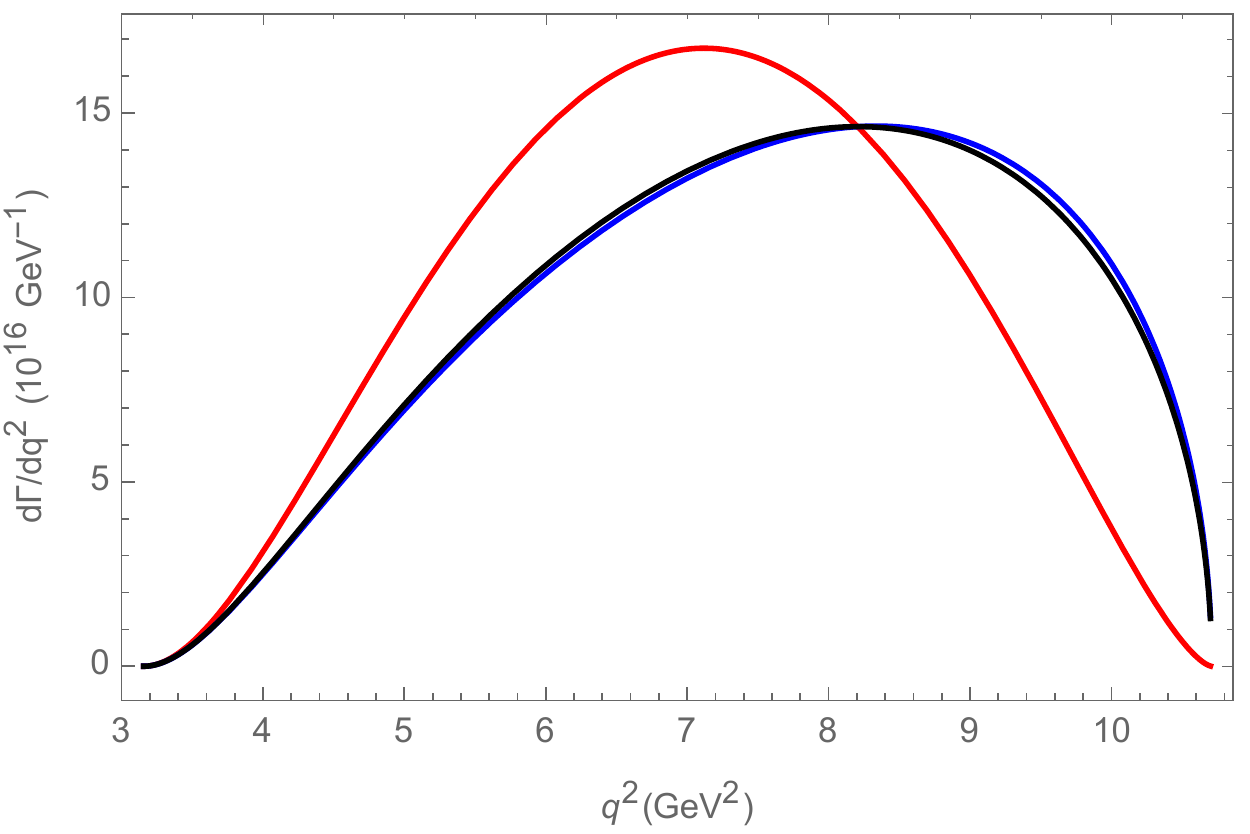}
\includegraphics[width=6cm]{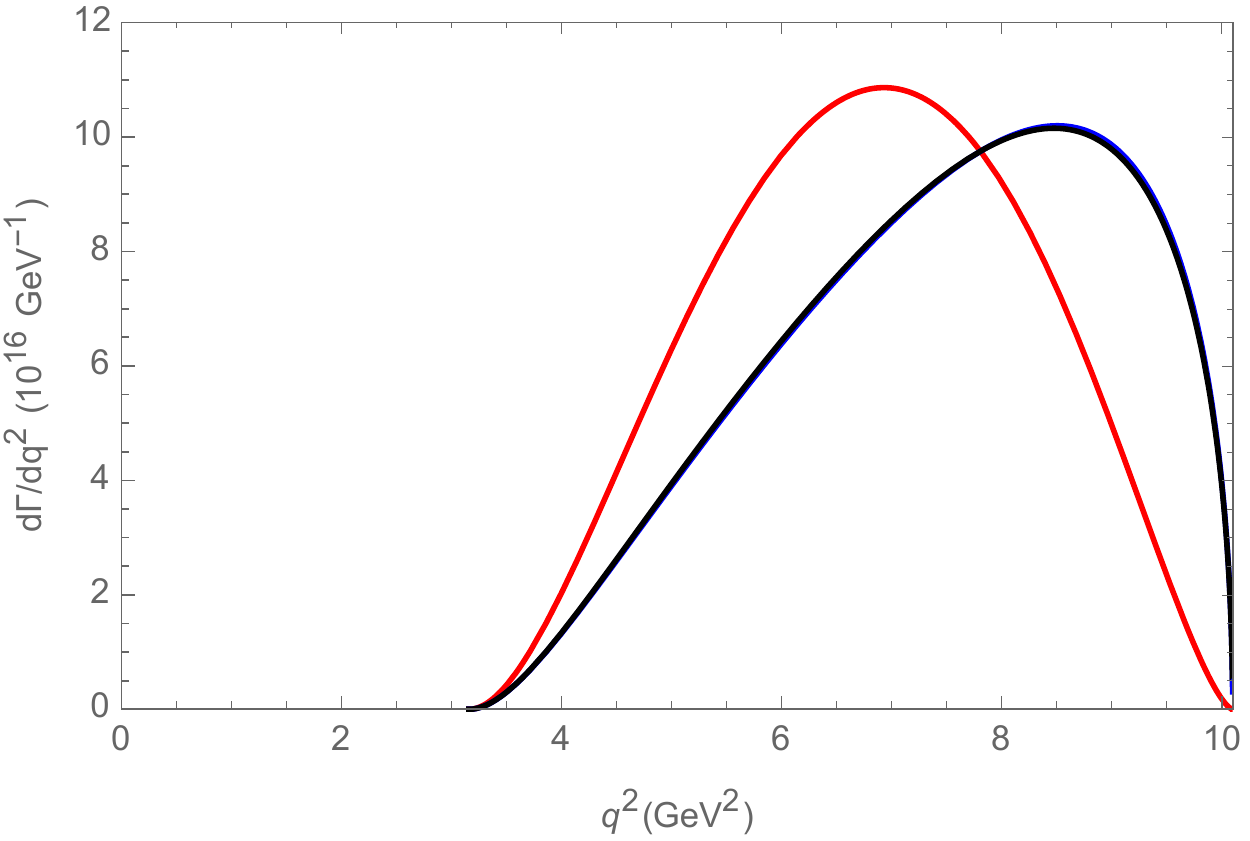}
\end{center}
\caption{Differential decay distribution $d\Gamma/dq^2$ for the SM, black curve; for a pure vector (axial-vector) hadronic form-factor, red (blue) curve normalized to the same total decay rate. Left panel for $B\to D^\star \tau \nu$ and right panel for $B_c\to J/\psi \tau \nu$.}
\label{f:dist}
\end{figure}

In the type of model we consider, in addition to the SM diagram, there is a $W^\prime$ mediated diagram as well as interference between these two. When all the neutrino masses can be neglected, there is no interference between the left and right-handed lepton currents and this allows us  to write simple formulas for both  ${d\Gamma}/{dq^2}$ and  the decay rate $\Gamma$ in terms of the corresponding SM results and the following two combinations of constants:
\begin{eqnarray}
F^{bc}_{\rm dir} &=& \left( 1+
\left(\frac{g_RM_W}{g_L M_{W^\prime}}\right)^4\frac{|V^\ell_{R3\ell}|^2|V_{Rcb}|^2}{{|V_{cb}|^2}}\right) \nonumber \\
F^{bc}_{{\rm mix}} &=& \xi_W\frac{g_R}{g_L}\frac{{\rm Re}\left(V_{cb}^\star V_{Rcb} \right)}{{|V_{cb}|^2}}\left(1-\left(\frac{M_W}{M_{W^\prime}}\right)^2\right)\left(1+\left(\frac{g_RM_W}{g_L M_{W^\prime}}\right)^2|V^\ell_{R3\ell}|^2\right).
\label{newfs}
\end{eqnarray}
The first term arises from the separate $W$ and $W^\prime$ contributions, whereas the second term is induced by  $W-W^\prime$ mixing. The superscript $bc$ is used to denote the $b\to c$ quark transition and is useful in order to generalize the notation to other cases. Our results for the different modes are then:

\begin{itemize}

\item Leptonic decay $B_c^\pm \to \ell^\pm \nu_\ell$. The hadronic transition in this case proceeds only through the axial vector form factor thus yielding
\begin{eqnarray}
\frac{\Gamma(B_c^-\to \tau^- \nu)}{\Gamma(B_c^-\to \tau^- \nu_\tau)_{SM}} &=&F^{bc}_{\rm dir}-2\ F^{bc}_{{\rm Mix}} .
\label{width}
\end{eqnarray}

\item Semileptonic decay $B\to D \tau \nu$. In this case only the vector form factor contributes to the hadronic transition resulting in
\begin{eqnarray}
\frac{d\Gamma(B\to D \tau \nu)}{dq^2} &=& \left.\frac{d\Gamma(B\to D \tau \nu)}{dq^2}\right|_{SM} \ \left(F^{bc}_{\rm dir}\ + 2\ F^{bc}_{\rm mix}\right)
\end{eqnarray}
This implies that the normalized distribution $1/\Gamma d\Gamma/dq^2$ for this mode is identical to the SM. Upon integration,
\begin{eqnarray}
R(D) = R(D)_{SM}\left(F^{bc}_{\rm dir}+2 F^{bc}_{\rm mix}\right).
\end{eqnarray}

\item Semileptonic decay $B\to D^\star \tau \nu$. This mode is more complicated in that both the vector and axial vector form factors contribute and they behave differently. Defining
\begin{eqnarray}
V = <D^\star|\bar{c}\gamma_\mu b|B>, && A = <D^\star|\bar{c}\gamma_\mu \gamma_5 b|B>
\end{eqnarray}
we can write the differential decay rate as
\begin{eqnarray}
\frac{d\Gamma(B\to D^\star \tau \nu)}{dq^2} &=& \left.\frac{d\Gamma(B\to D^\star \tau \nu)}{dq^2}\right|_{SM} \ \left(F^{bc}_{\rm dir}\ + 2\ F^{bc}_{\rm mix}\frac{|V|^2-|A|^2}{|V|^2+|A|^2}\right).
\label{diffds}
\end{eqnarray}

Using the values of the form factors given in Ref.~\cite{Lees:2013uzd}, the SM is dominated by the axial-vector hadronic form factor as can be seen from Figure~\ref{f:dist}. A transition through a purely axial-vector form factor (blue curve) has a spectrum shape almost indistinguishable from the SM case (black curve). On the other hand, a transition through a purely vector hadronic form factor produces a spectrum shifted towards lower $q^2$, as shown by the red curve in the same figure. The vector and axial-vector form factors do not interfere in this distribution, and their respective contributions to the total decay rate in the SM are
\begin{eqnarray}
\Gamma(B\to D^\star \tau \nu)=(0.41+6.92)\times 10^{-15}{\rm ~GeV}^{-1}=7.33\times 10^{-15}{\rm ~GeV}^{-1}.
\end{eqnarray}
This implies that the vector form factor contributes only $5.6\%$ of the SM rate. Integrating Eq.~\ref{diffds} we find
\begin{eqnarray}
R(D^\star) \approx R(D^\star)_{SM}\left(F^{bc}_{\rm dir}-1.77 F^{bc}_{\rm mix}\right),
\end{eqnarray}
very close to the result one would obtain from a pure axial-vector transition.

\item Semileptonic decay $B_c\to J/\psi \tau \nu$. This is also a pseudoscalar to vector transition so it has the same behaviour as the previous mode.
\begin{eqnarray}
\frac{d\Gamma(B_c\to J/\psi \tau \nu)}{dq^2} &=& \left.\frac{d\Gamma(B_c\to J/\psi \tau \nu)}{dq^2}\right|_{SM} \ \left(F^{bc}_{\rm dir}\ + 2\ F^{bc}_{\rm mix}\frac{|V'|^2-|A'|^2}{|V'|^2+|A'|^2}\right)
\end{eqnarray}
where now 
\begin{eqnarray}
V' = <J/\psi |\bar{c}\gamma_\mu b|B_c>, && A' = <J/\psi |\bar{c}\gamma_\mu \gamma_5 b|B_c>
\end{eqnarray}
In this case we use the values for the form factors as given in Ref.~\cite{Watanabe:2017mip} to display the SM differential distribution on the right panel of Figure~\ref{f:dist} as the black curve. Once again the blue (almost indistinguishable from the black one) and red curves illustrate transitions mediated by a purely axial-vector or vector form factors respectively, normalized to the total decay rate. We see that in this case the axial-vector hadronic form factor is even more dominant than in $B\to D^\star \tau \nu$. Their respective contributions to the total decay rate in the SM in this mode are 
\begin{eqnarray}
\Gamma(B_c\to J/\psi \tau \nu)=(0.067+4.27)\times 10^{-15}{\rm ~GeV}^{-1}=4.337\times 10^{-15}{\rm ~GeV}^{-1},
\end{eqnarray}
the vector form factor barely contributes  $1.5\%$ of the SM rate. It follows that for our model,
\begin{eqnarray}
R(J/\psi) \approx R(J/\psi)_{SM}\left(F^{bc}_{\rm dir}-1.94 F^{bc}_{\rm mix}\right).
\end{eqnarray}

\end{itemize}

\begin{figure}[!htb]
\begin{center}
\includegraphics[width=6cm]{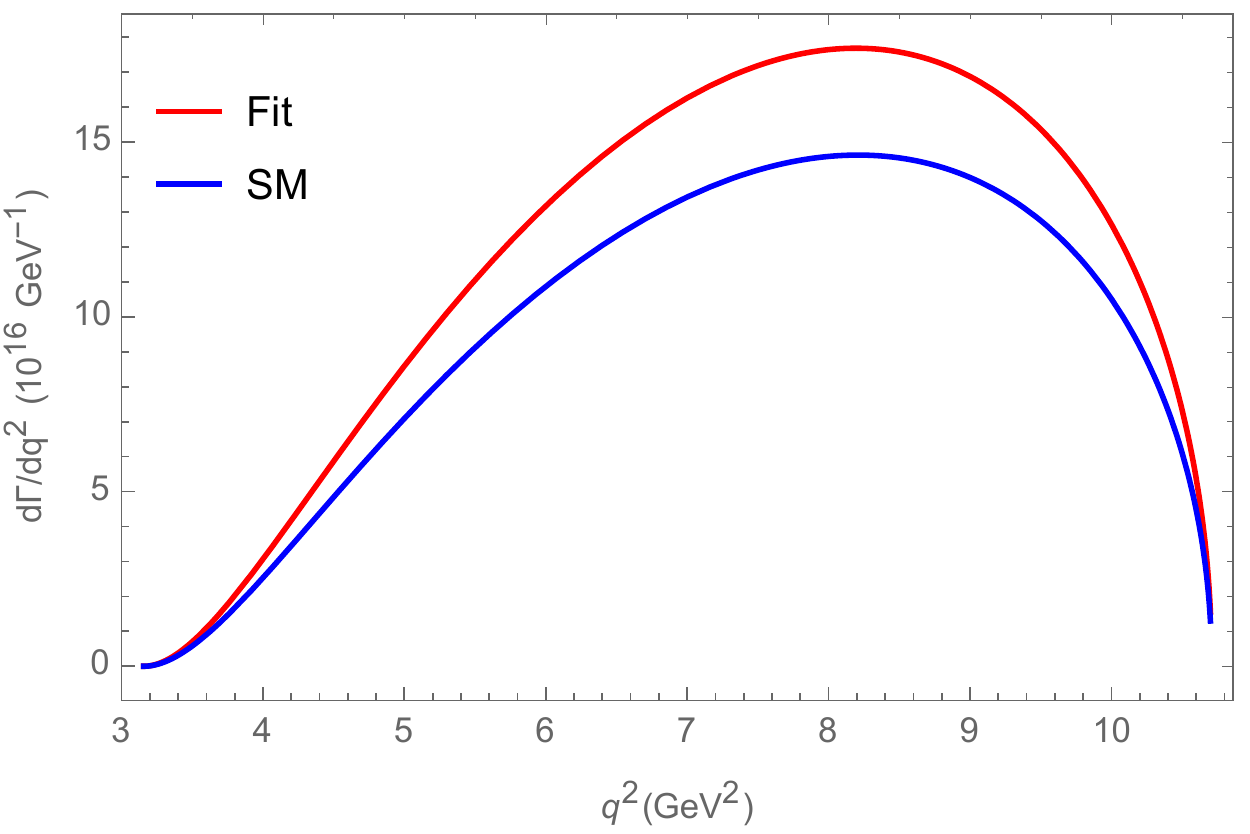}
\includegraphics[width=6cm]{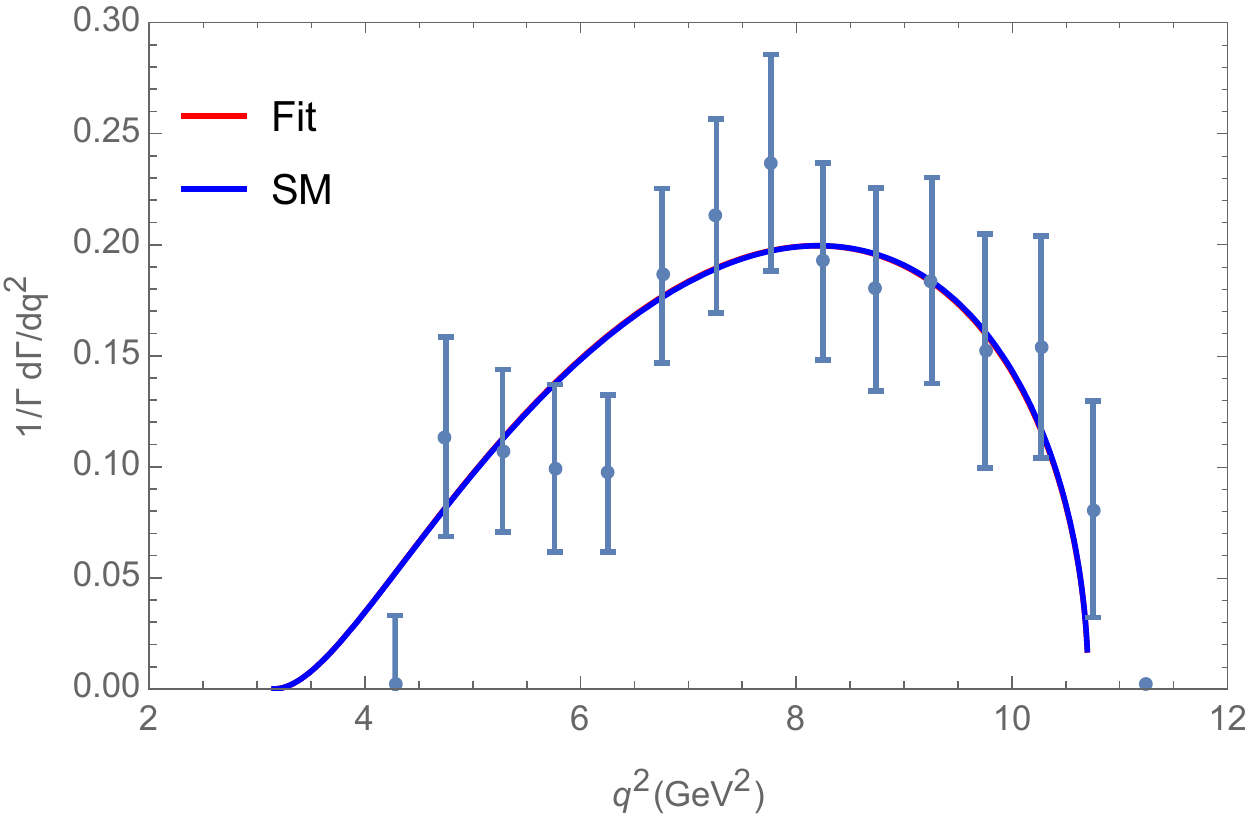}
\end{center}
\caption{Differential decay distribution $d\Gamma/dq^2$  in the SM and the  fit with the NP contributions as in Eq.~\ref{fitp} for $B\to D^\star \tau \nu$ (left panel). Normalized distributions compared to the BaBar data \cite{Lees:2013uzd} (right panel). Note how the model prediction for the shape of the distribution (red) is indistinguishable from the SM (blue) in the right panel because the mixing contribution is very small.}
\label{f:disr}
\end{figure}

The two experimental results for $R(D)$ and $R(D^\star)$ can be fit by the model choosing the parameters
\begin{eqnarray}
F^{bc}_{\rm dir} = 1.28,&& F^{bc}_{\rm mix}=0.04.
\label{fitp}
\end{eqnarray}
These parameters result in a differential distribution $d\Gamma/dq^2$ for $B\to D^\star \tau\nu$ that is dominated by the axial-form factor and is thus very similar to the SM one as shown in Figure~\ref{f:disr}. It should be clear from this result, in conjunction with the BaBar comparison of the SM vs observed distribution \cite{Lees:2013uzd}, that  $d\Gamma/dq^2$ is in agreement with the results of our model. This is shown in the right panel of Figure~\ref{f:disr} where the normalized distributions are compared with the BaBar data. The SM and our model are a good fit to the data and they are almost indistinguishable. This is due to the dominance of the $F^{bc}_{\rm dir}$ term over the mixing contribution.

\begin{figure}[!htb]
\begin{center}
\includegraphics[width=6cm]{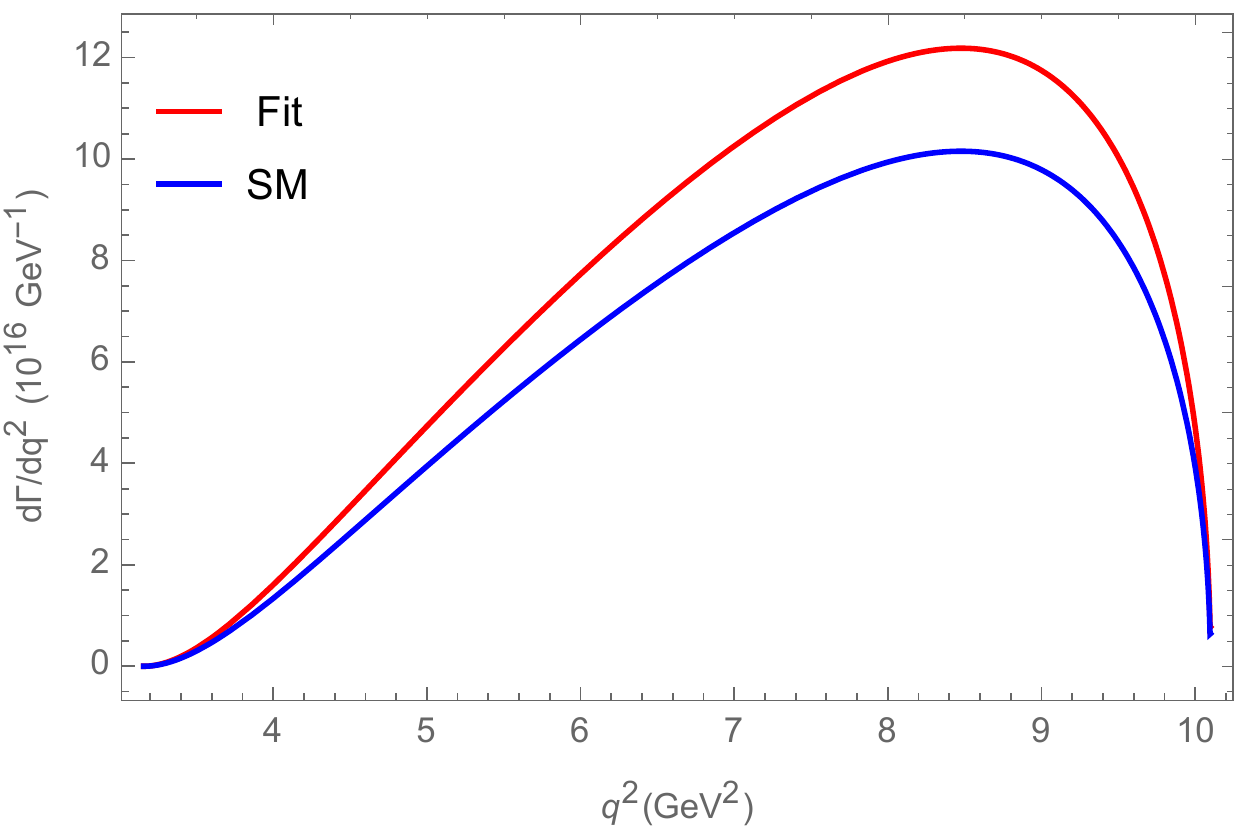}
\end{center}
\caption{Differential decay distribution $d\Gamma/dq^2$  in the SM and the best fit with the NP contributions as in Eq.~\ref{fitp} for  $B_c\to J/\psi \tau \nu$. }
\label{f:disrJ}
\end{figure}

Including the latest result, $R(J/\psi)$, does not change the fit significantly due to its large uncertainty. The prediction for this quantity, $R(J/\psi)= 0.34$, is thus on the low side of the central value by about 1.5 standard deviations. The differential distribution $d\Gamma/dq^2$ in this case is also very similar to the SM one, as seen in Figure~\ref{f:disrJ}, and would not serve to distinguish this model.

The rate $\Gamma(B_c^-\to \tau^- \nu)$ is predicted from Eq.~\ref{width} to be about 20\% larger than its SM value, which is well within the bound from the $B_c$ lifetime discussed in Ref.~\cite{Alonso:2016oyd}.

\section{Discussion}

We now examine the parameter values of Eq.~\ref{fitp} in the context of the model with RH currents as shown in Eq.~\ref{newfs}. The anomalies suggest that only the $\tau$ lepton is affected as the experiment sees no difference between muons and electrons. The model must then single out only the third family for enhancement with $|V^\ell_{R3\tau}|\sim 1$. Now $V_{Rcb}\equiv V^{u\star}_{Rtc}V^d_{Rbb}$ and $V^d_{Rbb}\sim 1$  and $V^{u\star}_{Rtc}$ is of the same order as $V_{cb}$, as discussed in our global analysis of the model in Ref.~\cite{He:2009ie}. This  requires $g_R /g_L\ M_W/M_{W^\prime} \sim 0.7$ to reproduce the first result of Eq.~\ref{newfs}. Requiring the model to be perturbative implies that $g_R/g_L \lsim  10$. If we take this ratio to be in the range $(5-10)$ we find in turn that $M_{W^\prime}$ is in the range $574-1150$~GeV, well within the direct reach of LHC.

The second combination of parameters, $F^{bc}_{\rm mix}$, requires that we re-examine bounds on $W-W^\prime$ mixing, in particular the combination $\xi_{eff}=\xi_W g_R/g_L$. This combination is constrained by $b\to s\gamma$~\cite{lrmodel,He:2002ha,Babu:1993hx} and we applied this constraint to our model in Ref.~\cite{He:2012zp}. We can update that result using the most recent HFLAV collaboration average: $B(b\to X_s \gamma) = (3.32 \pm 0.15)\times 10^{-4}$ \cite{Amhis:2016xyh} combined with the NNLL SM calculation  $B(b\to X_s \gamma) = (3.15 \pm 0.23)\times 10^{-4}$ \cite{Misiak:2006zs}. Assuming that the new physics interferes constructively with the SM, at the 3$\sigma$ level the allowed range becomes $-1.4 \times 10^{-3} \lsim \xi_{eff} \lsim 1.8 \times 10^{-3}$.

This range severely restricts the possible size of $F^{bc}_{{\rm mix}}$. For example, if we use $F^{bc}_{{\rm dir}}=1.28$ in Eq.~\ref{newfs} in combination with the above constraint on $\xi_{eff}$, 
we find
\begin{equation}
-2.1 \times 10^{-3} \lsim F^{bc}_{{\rm mix}} \lsim 2.7 \times 10^{-3},
\end{equation}
and it is not possible to reach the value $F^{bc}_{{\rm mix}}=0.04$ in the fit, Eq.~\ref{newfs}.
Under these conditions the three observables $R(D)$, $R(D^\star)$, and $R(J/\psi)$ become approximately proportional to the SM results,
\begin{eqnarray}
\frac{R(D)}{R(D)_{SM}}\approx \frac{R(D^\star)}{R(D^\star)_{SM}}\approx \frac{R(J/\psi)}{R(J/\psi)_{SM}}.
\end{eqnarray}
 It is interesting to notice that this universal enhancement of the three asymmetries reproduces what is found in models with an additional $SU(2)_L$ symmetry \cite{Boucenna:2016wpr,Boucenna:2016qad}. 
 
\begin{figure}[!htb]
\begin{center}
\includegraphics[width=8cm]{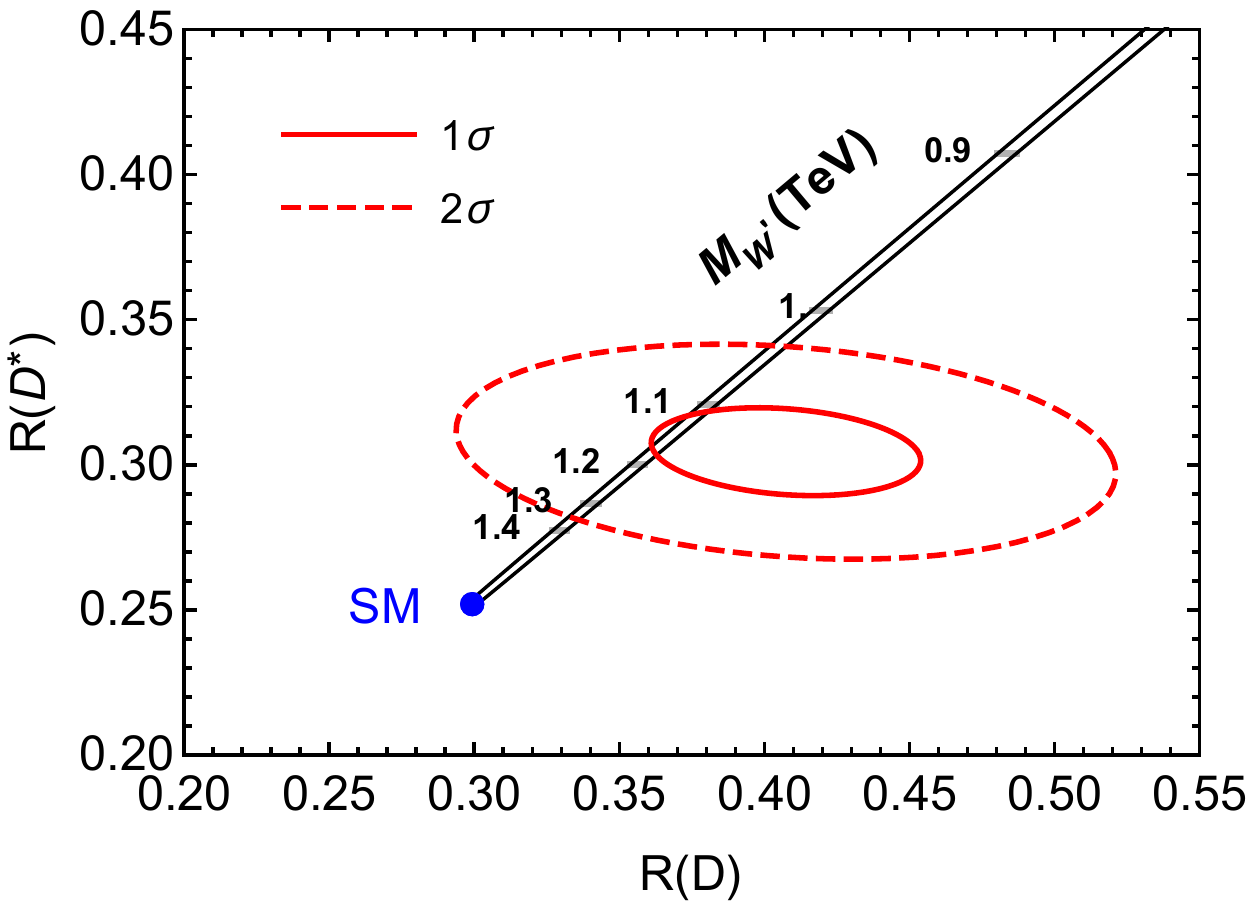}
\end{center}
\caption{$R(D)$ vs $R(D^\star)$. $1\sigma$ and $2\sigma$ allowed regions from the HFLAV collaboration \cite{Amhis:2016xyh} shown in red, the SM central values of Eq.~\ref{smpred} as the blue point and the predictions of this model as the black region. The tick marks along the model prediction indicate the required $W^\prime$ mass in TeV.}
\label{f:rdrds}
\end{figure}

We illustrate the situation in Figure~\ref{f:rdrds} which takes into account these constraints. The very narrow width of the model prediction range is due to the tight constraint on mixing and comparison with data implies that the model needs a $W^\prime$ mass very close to 1~TeV to successfully explain these anomalies.

The best direct limits on such a $W^\prime$ come from 19.7 fb$^{-1}$ of CMS data at $\sqrt{s}=8$~TeV \cite{Khachatryan:2015pua}. The first result presented in that paper excludes   an SSM $W^\prime$  \footnote{For SSM it is assumed that the $W^\prime$ is a heavy copy of the SM $W$, with its same couplings to fermions.} with mass below $2.7$~TeV. Since the production couplings  for  $W^\prime$ are model dependent it is more useful to quantify the constraint as $\sigma\times B(W^\prime \to \tau \nu) \lsim 3$~fb. 
In the SSM model considered by CMS $B(W^\prime \to \tau \nu)\approx 8.5\%$ for $W^\prime$ masses of order a TeV, where decay into top-bottom is allowed. In the non-universal model discussed here, this branching fraction approaches 25\% when $g_R>>g_L$ and $W^\prime$ couples almost exclusively to the third generation. 
 At the same time the production cross-section at LHC for our $W^\prime$ would be very suppressed due to its negligible couplings to the light fermions. Roughly then,
\begin{eqnarray}
\sigma\times B(W^\prime \to \tau \nu)\sim \left(\sigma\times B(W^\prime \to \tau \nu)\right)_{SM} \frac{25\%}{8.5\%}\left(\left|\frac{g_R}{g_L}\frac{V_{Rud}}{V_{ud}}\right|^2 {\rm ~or~}\xi_W^2\right).
\label{cross}
\end{eqnarray}
For the first term in the last bracket, corresponding to a direct coupling of the $W^\prime$ to the light quarks, we have: $V_{Rud}=V^{u\star}_{Rtu}V^d_{Rbd}$;  $V^d_{Rbd}\lsim 2.5\times 10^{-4}$ from $B_d$ mixing \cite{He:2006bk}; and fitting the existing body of FCNC constraints implies that $V^{u\star}_{Rtu} \sim 10^{-3}$ \cite{He:2009ie}. For the second term in the bracket we already saw that  $\xi_W$ is at most $10^{-3}$ in this scenario and we conclude that the corresponding $\sigma\times B(W^\prime \to \tau \nu)$ in our model is more than 6 orders of magnitude smaller than that of an SSM $W^\prime$ and the CMS data does not place any significant constraint.

The CMS paper also quantifies their result using a type of non-universal $W^\prime$ that also singles out the third generation  dubbed `NUGIM' \cite{Muller:1996dj,Chiang:2009kb}. In this case the CMS data excludes a $W^\prime$ with mass below $2.0-2.7$~TeV. Comparing the relevant figure of merit, $\sigma\times B(W^\prime \to \tau \nu)$, of this model to ours we see that $B(W^\prime \to \tau \nu)$ can be quite similar but 
\begin{eqnarray}
\sigma(pp\to W^\prime)_{\rm this~model}\sim \sigma(pp\to W^\prime)_{\rm NUGIM} \left|\frac{V_{Rud}}{(s_E/c_E)}\right|^2 .
\label{crossh}
\end{eqnarray}
Whereas $V_{Rud}$ can (and in fact is constrained to be) very small, the parameter $s_E/c_E$ of NUGIM is of order one (for this reason the $W-W^\prime$ mixing in the NUGIM model is not important in $\sigma(pp\to W^\prime)$). The net result is that the CMS limits do not directly apply to our model. A separate study is needed for an accurate comparison of our model to LHC results, taking into account production from heavier quarks.

As mentioned before, our model relies on the existence of an additional light neutrino to explain these anomalies and this can have other observable consequences. In Ref.~\cite{He:2012zp} we have already seen that there are no significant constraints from the invisible $Z$ width. At the same time the model can provide an {\it enhancement} to the rare $K\to \pi \nu \nu$ modes \cite{He:2004it} where new results are expected from NA62 and KOTO. 

The existence of a light right-handed neutrino contributes to the effective neutrino number $\Delta N_{eff}$ which is also constrained by  cosmological considerations and this may affect the viability of our model. There is some uncertainty as to the value of this constraint, 
but commonly used numbers are, for example \cite{effective-neutrino},
\begin{eqnarray}
\Delta N_{eff} < \begin{cases} 0.28 &\mbox{for } H_0 = 68.7^{+0.6}_{-0.7}{\rm ~km/s/Mpc} \\ 
0.77 & \mbox{for } H_0 = 71.3^{+1.9}_{-2.2}{\rm ~km/s/Mpc}  \end{cases} 
\end{eqnarray}
As we saw above, our model requires 
\begin{eqnarray}
\left(\frac{g_RM_W}{g_L M_{W^\prime}}\right)^4\frac{|V^\ell_{R3\tau}|^2|V_{Rcb}|^2}{{|V_{cb}|^2}} \sim 0.3
\end{eqnarray}
to explain the $R_{D^{(*)}}$ anomalies and this is only slightly weaker than the usual weak interaction.  At the same time, the exchange of a $W^\prime$ can bring the new $\nu_R$ into thermal equilibrium with the SM particles through scattering of right-handed neutrinos with tauons at a rate proportional to $(g_RM_W/g_L M_{W'})^4|V^\ell_{R3\tau}|^4$ relative to the usual weak interaction. In fact, with $V_{Rcb}/V_{cb} \sim 1$ this would result in $\Delta N_{eff} \sim 1$ bringing into question the viability of our model.  The mixing induced interaction, proportional to $\xi_W$, is smaller and does not lead to large contributions to $\Delta N_{eff}$.

However, the aforementioned scattering  of right-handed neutrinos with tauons  is only effective for temperatures $T_R$ above $T_{\tau} \sim m_\tau$. At the time of big bang nucleosynthesis (BBN), the temperature is about $T_{BBN} \sim 1$~MeV, implying that $\Delta N_{eff}$ is suppressed by a factor 
\begin{eqnarray}
r=\left(\frac{g_*(T_{BBN})}{g_*(T_R)}\right)^{4/3}
\end{eqnarray}
where $g_*(T)$ is the effective number of relativistic degrees of freedom at temperature $T$ and $g_*(T_{BBN}) = 10.75$. In addition, 
$g_*(T_R)$ is larger than $g_*(T_{QCD}) \sim 58$ since the QCD phase transition temperature $T_{QCD}$ is of order a few hundred MeV \cite{dolgov}. All this implies that the contribution to $\Delta N_{eff}$ from our additional neutrino is less than 0.1 and safely within the BBN constraint. 

Similarly, $\tau$ decay processes into $\nu_R$ plus other SM particles are also suppressed by the same factor $r$, but one might worry about additional processes without this suppression. For example,  $\nu_R$ scattering off an electron or a muon. However, these are proportional to the additional mixing parameters $|V^\ell_{R3e(\mu)}|^4$ and can be made sufficiently small by lowering $V^\ell_{R3e(\mu)}$. 

Another potentially worrisome process is the exchange of a $Z'$ in the scattering of a $\nu_R$  off an electron or SM neutrino $\nu_L$. In this case the interaction strength is proportional to $(g^2_Y/M_{Z'}^2)^2$ \cite{He:2003qv}, and when compared to $Z$ exchange induced $\nu_L$ scattering off an electron or $\nu_L$, it is suppressed by a factor of $(M_Z/M_{Z'})^4$. The constraint on $\Delta N_{eff}$ becomes in this case a lower bound on the $Z^\prime$ mass, $M_{Z'}\gsim 200$~GeV.

In conclusion we find that new right handed currents affect the semi-tauonic $B$ decay anomalies in a way that is consistent with current bounds, including those on the effective number of neutrino species from BBN. A confirmation of a high value for $R(J/\psi)$ would exclude them as a viable explanation and would also exclude new left-handed currents. The most promising way to rule out this explanation of the anomalies is the exclusion of a $W^\prime$ in the $\tau$-channel at LHC in the mass range $1-1.4$~TeV. The suppression of our $W^\prime$ couplings to light fermions significantly complicates this comparison.

\begin{acknowledgments}

\end{acknowledgments}

We thank Avelino Vicente for clarifications on Refs.~\cite{Boucenna:2016wpr,Boucenna:2016qad}. 
X.G.H. was supported in part by MOST of ROC (Grant No. MOST104-2112-M-002-015-MY3) and in part by NSFC of PRC (Grant Nos. 11575111 and 11735010), and partially supported by a grant from Science and Technology Commission of Shanghai Municipality (Grants No.16DZ2260200) and National Natural Science Foundation of China (Grants No.11655002).


\begin{thebibliography}{999}
   
\bibitem{Lees:2012xj} 
  J.~P.~Lees {\it et al.} [BaBar Collaboration],
  Phys.\ Rev.\ Lett.\  {\bf 109}, 101802 (2012)
  doi:10.1103/PhysRevLett.109.101802
  [arXiv:1205.5442 [hep-ex]].
  
\bibitem{Lees:2013uzd} 
  J.~P.~Lees {\it et al.} [BaBar Collaboration],
  Phys.\ Rev.\ D {\bf 88}, no. 7, 072012 (2013)
  doi:10.1103/PhysRevD.88.072012
  [arXiv:1303.0571 [hep-ex]].
  
\bibitem{Huschle:2015rga} 
  M.~Huschle {\it et al.} [Belle Collaboration],
  Phys.\ Rev.\ D {\bf 92}, no. 7, 072014 (2015)
  doi:10.1103/PhysRevD.92.072014
  [arXiv:1507.03233 [hep-ex]].
  
\bibitem{Sato:2016svk} 
  Y.~Sato {\it et al.} [Belle Collaboration],
  Phys.\ Rev.\ D {\bf 94}, no. 7, 072007 (2016)
  doi:10.1103/PhysRevD.94.072007
  [arXiv:1607.07923 [hep-ex]].
   
\bibitem{Hirose:2016wfn} 
  S.~Hirose {\it et al.} [Belle Collaboration],
  Phys.\ Rev.\ Lett.\  {\bf 118}, no. 21, 211801 (2017)
  doi:10.1103/PhysRevLett.118.211801
  [arXiv:1612.00529 [hep-ex]].
 
  
\bibitem{Aaij:2015yra} 
  R.~Aaij {\it et al.} [LHCb Collaboration],
  Phys.\ Rev.\ Lett.\  {\bf 115}, no. 11, 111803 (2015)
  Erratum: [Phys.\ Rev.\ Lett.\  {\bf 115}, no. 15, 159901 (2015)]
  doi:10.1103/PhysRevLett.115.159901, 10.1103/PhysRevLett.115.111803
  [arXiv:1506.08614 [hep-ex]].
  
 \bibitem{LHCbconf}
 LHCb Collaboration at FPCP2017.
 
\bibitem{Amhis:2016xyh} 
http://www.slac.stanford.edu/xorg/hflav/, 
  Y.~Amhis {\it et al.},
  arXiv:1612.07233 [hep-ex].
 
\bibitem{Lattice:2015rga} 
  J.~A.~Bailey {\it et al.} [MILC Collaboration],
  Phys.\ Rev.\ D {\bf 92}, no. 3, 034506 (2015)
  doi:10.1103/PhysRevD.92.034506
  [arXiv:1503.07237 [hep-lat]].
  
\bibitem{Na:2015kha} 
  H.~Na {\it et al.} [HPQCD Collaboration],
  Phys.\ Rev.\ D {\bf 92}, no. 5, 054510 (2015)
  Erratum: [Phys.\ Rev.\ D {\bf 93}, no. 11, 119906 (2016)]
  doi:10.1103/PhysRevD.93.119906, 10.1103/PhysRevD.92.054510
  [arXiv:1505.03925 [hep-lat]].
  
\bibitem{Fajfer:2012vx} 
  S.~Fajfer, J.~F.~Kamenik and I.~Nisandzic,
  Phys.\ Rev.\ D {\bf 85}, 094025 (2012)
  doi:10.1103/PhysRevD.85.094025
  [arXiv:1203.2654 [hep-ph]].
  
\bibitem{Aaij:2017tyk} 
  R.~Aaij {\it et al.} [LHCb Collaboration],
  arXiv:1711.05623 [hep-ex].
  

\bibitem{Anisimov:1998uk} 
  A.~Y.~Anisimov, I.~M.~Narodetsky, C.~Semay and B.~Silvestre-Brac,
  Phys.\ Lett.\ B {\bf 452}, 129 (1999)
  doi:10.1016/S0370-2693(99)00273-7
  [hep-ph/9812514].
  
\bibitem{Kiselev:2002vz} 
  V.~V.~Kiselev,
  hep-ph/0211021.
  
\bibitem{Ivanov:2006ni} 
  M.~A.~Ivanov, J.~G.~Korner and P.~Santorelli,
  Phys.\ Rev.\ D {\bf 73}, 054024 (2006)
  doi:10.1103/PhysRevD.73.054024
  [hep-ph/0602050].
  
\bibitem{Hernandez:2006gt} 
  E.~Hernandez, J.~Nieves and J.~M.~Verde-Velasco,
  Phys.\ Rev.\ D {\bf 74}, 074008 (2006)
  doi:10.1103/PhysRevD.74.074008
  [hep-ph/0607150].
  
\bibitem{Watanabe:2017mip} 
  R.~Watanabe,
  arXiv:1709.08644 [hep-ph].
 
\bibitem{Kamenik:2008tj} 
  J.~F.~Kamenik and F.~Mescia,
  Phys.\ Rev.\ D {\bf 78}, 014003 (2008)
  [arXiv:0802.3790 [hep-ph]].
 
 
\bibitem{Tanaka:2010se} 
  M.~Tanaka and R.~Watanabe,
  Phys.\ Rev.\ D {\bf 82}, 034027 (2010)
  [arXiv:1005.4306 [hep-ph]].

\bibitem{Crivellin:2012ye} 
  A.~Crivellin, C.~Greub and A.~Kokulu,
  arXiv:1206.2634 [hep-ph].
 
\bibitem{Celis:2012dk} 
  A.~Celis, M.~Jung, X.~Q.~Li and A.~Pich,
  JHEP {\bf 1301}, 054 (2013)
  doi:10.1007/JHEP01(2013)054
  [arXiv:1210.8443 [hep-ph]].
  
\bibitem{Deshpande:2012rr} 
  N.~G.~Deshpande and A.~Menon,
  arXiv:1208.4134 [hep-ph].
  

\bibitem{Fajfer:2012jt} 
  S.~Fajfer, J.~F.~Kamenik, I.~Nisandzic and J.~Zupan,
  arXiv:1206.1872 [hep-ph]; 
  
\bibitem{Datta:2012qk} 
  A.~Datta, M.~Duraisamy and D.~Ghosh,
  Phys.\ Rev.\ D {\bf 86}, 034027 (2012)
  [arXiv:1206.3760 [hep-ph]];
  
\bibitem{Becirevic:2012jf} 
  D.~Becirevic, N.~Kosnik and A.~Tayduganov,
  Phys.\ Lett.\ B {\bf 716}, 208 (2012)
  [arXiv:1206.4977 [hep-ph]].
  
\bibitem{He:2012zp} 
  X.~G.~He and G.~Valencia,
  Phys.\ Rev.\ D {\bf 87}, no. 1, 014014 (2013)
  doi:10.1103/PhysRevD.87.014014
  [arXiv:1211.0348 [hep-ph]].
 
\bibitem{Rashed:2012bd} 
  A.~Rashed, M.~Duraisamy and A.~Datta,
  Phys.\ Rev.\ D {\bf 87}, no. 1, 013002 (2013)
  doi:10.1103/PhysRevD.87.013002
  [arXiv:1204.2023 [hep-ph]].
  
\bibitem{Sakaki:2012ft} 
  Y.~Sakaki and H.~Tanaka,
  Phys.\ Rev.\ D {\bf 87}, no. 5, 054002 (2013)
  doi:10.1103/PhysRevD.87.054002
  [arXiv:1205.4908 [hep-ph]].
  
\bibitem{Tanaka:2012nw} 
  M.~Tanaka and R.~Watanabe,
  Phys.\ Rev.\ D {\bf 87}, no. 3, 034028 (2013)
  doi:10.1103/PhysRevD.87.034028
  [arXiv:1212.1878 [hep-ph]].
  
\bibitem{Ko:2012sv} 
  P.~Ko, Y.~Omura and C.~Yu,
  JHEP {\bf 1303}, 151 (2013)
  doi:10.1007/JHEP03(2013)151
  [arXiv:1212.4607 [hep-ph]].
  
  
\bibitem{Bambhaniya:2013wza} 
  G.~Bambhaniya, J.~Chakrabortty, J.~Gluza, M.~Kordiaczy?ska and R.~Szafron,
  JHEP {\bf 1405}, 033 (2014)
  doi:10.1007/JHEP05(2014)033
  [arXiv:1311.4144 [hep-ph]].
  
\bibitem{Atoui:2013zza} 
  M.~Atoui, V.~MorŽnas, D.~Be?irevic and F.~Sanfilippo,
  Eur.\ Phys.\ J.\ C {\bf 74}, no. 5, 2861 (2014)
  doi:10.1140/epjc/s10052-014-2861-z
  [arXiv:1310.5238 [hep-lat]].
 
\bibitem{Dutta:2013qaa} 
  R.~Dutta, A.~Bhol and A.~K.~Giri,
  Phys.\ Rev.\ D {\bf 88}, no. 11, 114023 (2013)
  doi:10.1103/PhysRevD.88.114023
  [arXiv:1307.6653 [hep-ph]].
  
\bibitem{Freytsis:2015qca} 
  M.~Freytsis, Z.~Ligeti and J.~T.~Ruderman,
  Phys.\ Rev.\ D {\bf 92}, no. 5, 054018 (2015)
  doi:10.1103/PhysRevD.92.054018
  [arXiv:1506.08896 [hep-ph]].
  
\bibitem{Boucenna:2016wpr} 
  S.~M.~Boucenna, A.~Celis, J.~Fuentes-Martin, A.~Vicente and J.~Virto,
  Phys.\ Lett.\ B {\bf 760}, 214 (2016)
  doi:10.1016/j.physletb.2016.06.067
  [arXiv:1604.03088 [hep-ph]].
   
\bibitem{Boucenna:2016qad} 
  S.~M.~Boucenna, A.~Celis, J.~Fuentes-Martin, A.~Vicente and J.~Virto,
  JHEP {\bf 1612}, 059 (2016)
  doi:10.1007/JHEP12(2016)059
  [arXiv:1608.01349 [hep-ph]].
  
\bibitem{Chiang:2016qov} 
  C.~W.~Chiang, X.~G.~He and G.~Valencia,
  Phys.\ Rev.\ D {\bf 93}, no. 7, 074003 (2016)
  doi:10.1103/PhysRevD.93.074003
  [arXiv:1601.07328 [hep-ph]].

\bibitem{Zhu:2016xdg} 
  J.~Zhu, H.~M.~Gan, R.~M.~Wang, Y.~Y.~Fan, Q.~Chang and Y.~G.~Xu,
  Phys.\ Rev.\ D {\bf 93}, no. 9, 094023 (2016)
  doi:10.1103/PhysRevD.93.094023
  [arXiv:1602.06491 [hep-ph]].
  
\bibitem{Kim:2016yth} 
  C.~S.~Kim, G.~Lopez-Castro, S.~L.~Tostado and A.~Vicente,
  Phys.\ Rev.\ D {\bf 95}, no. 1, 013003 (2017)
  doi:10.1103/PhysRevD.95.013003
  [arXiv:1610.04190 [hep-ph]].

\bibitem{Celis:2016azn} 
  A.~Celis, M.~Jung, X.~Q.~Li and A.~Pich,
  Phys.\ Lett.\ B {\bf 771}, 168 (2017)
  doi:10.1016/j.physletb.2017.05.037
  [arXiv:1612.07757 [hep-ph]].
 
\bibitem{Dutta:2017xmj} 
  R.~Dutta and A.~Bhol,
  Phys.\ Rev.\ D {\bf 96}, no. 7, 076001 (2017)
  doi:10.1103/PhysRevD.96.076001
  [arXiv:1701.08598 [hep-ph]].
 
\bibitem{Cvetic:2017gkt} 
  G.~Cveti?, F.~Halzen, C.~S.~Kim and S.~Oh,
  Chin.\ Phys.\ C {\bf 41}, no. 11, 113102 (2017)
  doi:10.1088/1674-1137/41/11/113102
  [arXiv:1702.04335 [hep-ph]].
 
\bibitem{Ko:2017lzd} 
  P.~Ko, Y.~Omura, Y.~Shigekami and C.~Yu,
  Phys.\ Rev.\ D {\bf 95}, no. 11, 115040 (2017)
  doi:10.1103/PhysRevD.95.115040
  [arXiv:1702.08666 [hep-ph]].
 
\bibitem{Chen:2017hir} 
  C.~H.~Chen, T.~Nomura and H.~Okada,
  Phys.\ Lett.\ B {\bf 774}, 456 (2017)
  doi:10.1016/j.physletb.2017.10.005
  [arXiv:1703.03251 [hep-ph]].
  
\bibitem{Chen:2017eby} 
  C.~H.~Chen and T.~Nomura,
  Eur.\ Phys.\ J.\ C {\bf 77}, no. 9, 631 (2017)
  doi:10.1140/epjc/s10052-017-5198-6
  [arXiv:1703.03646 [hep-ph]].
  
\bibitem{Dutta:2017wpq} 
  R.~Dutta,
  arXiv:1710.00351 [hep-ph].
  
  

  
\bibitem{Faustov:2012nk} 
  R.~N.~Faustov and V.~O.~Galkin,
  arXiv:1207.5973 [hep-ph].
  

\bibitem{He:2002ha}
  X.~G.~He and G.~Valencia,
  Phys.\ Rev.\  D {\bf 66}, 013004 (2002)
  [Erratum-ibid.\  D {\bf 66}, 079901 (2002)]
  [arXiv:hep-ph/0203036];
  
\bibitem{He:2003qv}
  X.~G.~He and G.~Valencia,
  Phys.\ Rev.\  D {\bf 68}, 033011 (2003)
  [arXiv:hep-ph/0304215].

\bibitem{Alonso:2016oyd} 
  R.~Alonso, B.~Grinstein and J.~Martin Camalich,
  Phys.\ Rev.\ Lett.\  {\bf 118}, no. 8, 081802 (2017)
  doi:10.1103/PhysRevLett.118.081802
  [arXiv:1611.06676 [hep-ph]].
  
  
\bibitem{He:2009ie} 
  X.~-G.~He and G.~Valencia,
  Phys.\ Lett.\ B {\bf 680}, 72 (2009)
  [arXiv:0907.4034 [hep-ph]].
  
 \bibitem{lrmodel} R.N. Mohapatra and J.C. Pati, Phys. Rev. {\bf D11},
566(1975); Phys. Rev. {\bf D11}, 2558(1975);
R.N. Mohapatra and G. Senjanovic, Phys. Rev. {\bf D12}, 1502(1975);
P.~Langacker and S.~Uma Sankar,
Phys.\ Rev.\ D {\bf 40}, 1569 (1989).
 
  
\bibitem{Babu:1993hx} 
  K.~S.~Babu, K.~Fujikawa and A.~Yamada,
  Phys.\ Lett.\ B {\bf 333}, 196 (1994)
  [hep-ph/9312315].

\bibitem{Misiak:2006zs} 
  M.~Misiak, H.~M.~Asatrian, K.~Bieri, M.~Czakon, A.~Czarnecki, T.~Ewerth, A.~Ferroglia and P.~Gambino {\it et al.},
  Phys.\ Rev.\ Lett.\  {\bf 98}, 022002 (2007)
  [hep-ph/0609232].

  
\bibitem{Khachatryan:2015pua} 
  V.~Khachatryan {\it et al.} [CMS Collaboration],
  Phys.\ Lett.\ B {\bf 755}, 196 (2016)
  doi:10.1016/j.physletb.2016.02.002
  [arXiv:1508.04308 [hep-ex]].
  
\bibitem{He:2006bk} 
  X.~G.~He and G.~Valencia,
  Phys.\ Rev.\ D {\bf 74}, 013011 (2006)
  doi:10.1103/PhysRevD.74.013011
  [hep-ph/0605202].
  
 
  
\bibitem{Muller:1996dj} 
  D.~J.~Muller and S.~Nandi,
  Phys.\ Lett.\ B {\bf 383}, 345 (1996)
  doi:10.1016/0370-2693(96)00745-9
  [hep-ph/9602390].
  
\bibitem{Chiang:2009kb} 
  C.~W.~Chiang, N.~G.~Deshpande, X.~G.~He and J.~Jiang,
  Phys.\ Rev.\ D {\bf 81}, 015006 (2010)
  doi:10.1103/PhysRevD.81.015006
  [arXiv:0911.1480 [hep-ph]].

\bibitem{He:2004it} 
  X.~G.~He and G.~Valencia,
  Phys.\ Rev.\ D {\bf 70}, 053003 (2004)
  doi:10.1103/PhysRevD.70.053003
  [hep-ph/0404229].
  
  \bibitem{effective-neutrino}
  J.~L.~Bernal, L.~Verde and A.~G.~Riess,
  JCAP {\bf 1610}, no. 10, 019 (2016)
  doi:10.1088/1475-7516/2016/10/019
  [arXiv:1607.05617 [astro-ph.CO]].
  
  \bibitem{dolgov}
  A.~D.~Dolgov,
  Phys.\ Rept.\  {\bf 370}, 333 (2002)
  doi:10.1016/S0370-1573(02)00139-4
  [hep-ph/0202122].
     
\end{thebibliography}
\end{document}